# Surface imaging of cool evolved stars in the era of the ELT

M. Wittkowski[1]
A. Chiavassa[2]
S. Höfner[3]
J. B. Climent[4]

[1] ESO
[2] Université Côte d'Azur, Nice, France
[3] Uppsala University, Sweden
[4] Universitat de València

**Cool evolved stars are the main source of chemical enrichment of the interstellar medium. Understanding their mass loss offers a unique opportunity to study the cycle of matter. We discuss interferometric studies and their comparison to latest state-of-the-art dynamic model atmospheres. They show broad agreement for asymptotic giant branch stars. For red supergiants, however, current models cannot explain observed extensions by far, pointing to missing physical processes in their models, and uncertainties in our general understanding of mass loss. We present ongoing imaging and time-series observations that may provide the strongest constraint and may help to identify missing dynamic processes. VLTI studies will remain the highest spatial resolution observations at ESO into the ELT era, complemented by ALMA observations. We discuss crucial improvements in both instrumental and operational areas for surface imaging of cool evolved stars in the era of the ELT.**

## Introduction

Low-to-intermediate mass asymptotic giant branch (AGB) stars and massive red supergiants (RSGs) are cool evolved stars with low effective temperatures between about 2500 K and 4000 K, and together spanning a huge range of luminosities of log $L/L_{sun}$ between about 3 and 6. They experience a mass loss rate of up to about $10^{-5}$ to $10^{-4}$ $M_{sun}$/year, precipitating the return of material to the interstellar medium. The mass-loss process is thought to be triggered by a levitation of the atmosphere to radii where dust can form, and radiative acceleration on dust grains dragging along the gas (e.g., Höfner and Olofsson 2018). However, some details of this process are surprisingly little understood, in particular for red supergiants, for which current models of pulsation and convection alone cannot elevate the atmosphere to observed extensions (Arroyo-Torres et al. 2015). Other processes may need to be included, such as radiative acceleration on Doppler-shifted molecular lines (Josselin & Plez 2007), magnetic fields and Alfén waves (e.g., Thirumalai & Heyl 2012, Airapetian et al. 2010), rotation (e.g., Vlemmings et al. 2018), or giant hot spots (e.g., Montarges et al. 2016, Vlemmings et al. 2017). The mass-loss is initiated on top of the convective photosphere of these sources, a crucial region that we need to image and study for a representative sample covering the large parameter space, to understand the physical mechanisms that elevate the atmosphere and drive the mass loss, by comparison to theoretical simulations.

## Current stage

A decade of comparisons of mostly single-epoch spectro-interferometry at individual baselines to dynamic model atmospheres for AGB and RSG stars showed that both classes of evolved stars observationally show similarly extended molecular layers that can be fairly well reproduced by 1D pulsation dynamic models and with 3D models of convection for AGB stars (e.g., Wittkowski et al. 2016), but that by far cannot be reproduced by current 1D or 3D dynamic models of RSG stars (Arroyo-Torres et al. 2015).

Interferometry now succeeds in imaging stellar surfaces and their environments, allowing us to compare reconstructed images and 3D simulations of convection in terms of contrast and morphology. Figs. 1-3 show as examples our work based on the PIONIER instrument in the near-IR H band for the the carbon AGB star R Scl (Wittkowski et al. 2017a), and the RSGs V766 Cen (Wittkowski et al. 2017b) and V602 Car (Climent et al., in prep.). All image reconstructions show a complex structure within the stellar disk. The image reconstruction of R Scl shows a complex structure and a dominant spot with large contrast that cannot be explained by structure on the stellar surface itself, but that we interpret as the effects of convection on clumpy dust formation at 2-3 stellar radii as outlined on the right part of the figure. The surface structure of the RSGs shows contrasts of about 10% as predicted by 3D simulations of convection at the photospheric layer of RSGs. The morphology of the arc-like surface feature of V602 Car can be reproduced by some episodes of 3D simulations of convection as shown by the comparison in Fig. 3. Despite this agreement at the photospheric layer, the need for an additional extended component is still present to describe the visibility data for this source.

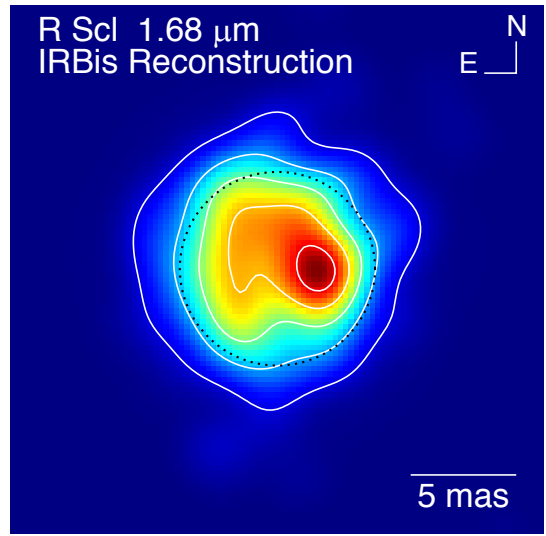

Figure 1a. Image reconstruction of the carbon AGB star R Scl based on VLTI-PIONIER data showing a dominant spot of large contrast. From Wittkowski et al. (2017a).

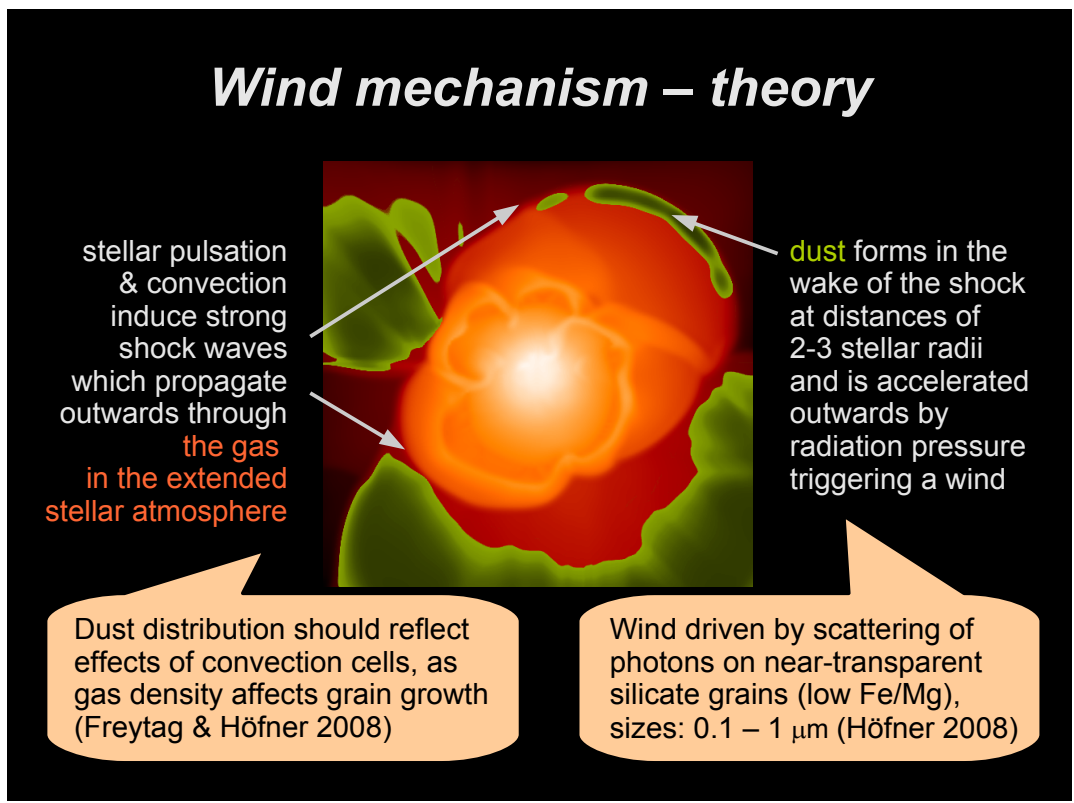

Figure 1b. Interpretation of the dominant spot of Fig. 1a as the effect of convection on clumpy dust formation at 2-3 stellar radii. Based on Freytag & Höfner (2008), Höfner (2008), Freytag et al. (2017).

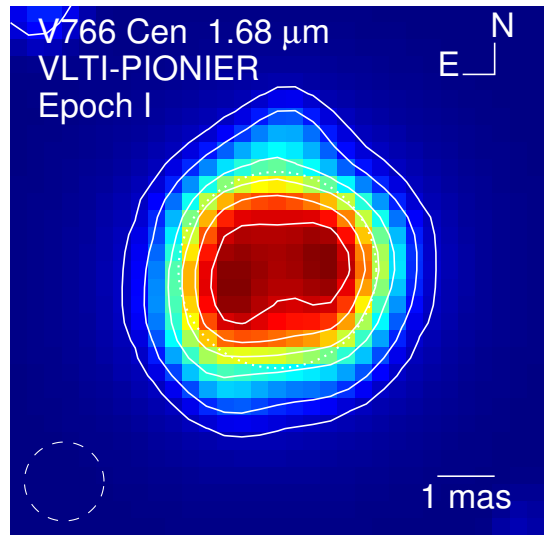

Figure 2. Image reconstruction of the RSG/hypergiant V766 Cen based on VLTI-PIONIER data. From Wittkowski et al. (2017b).

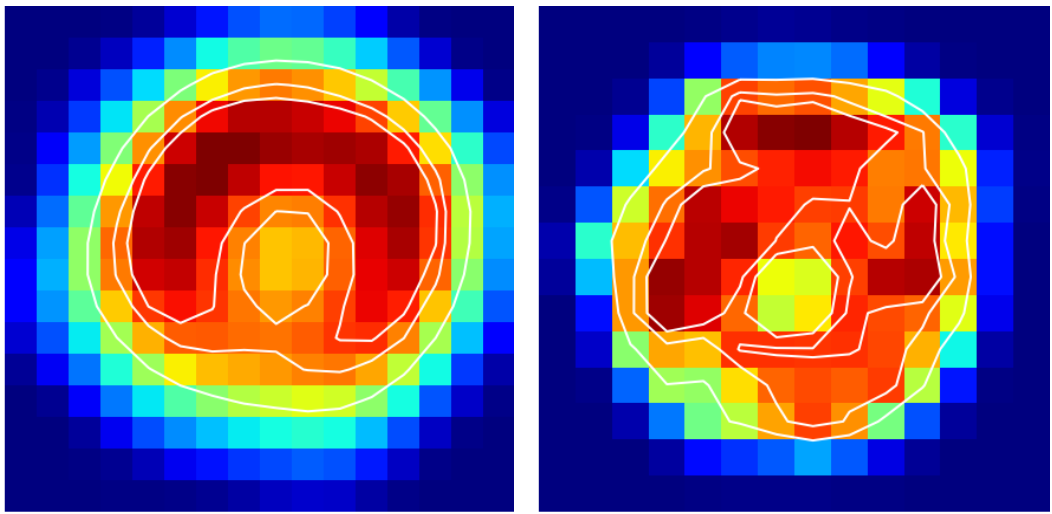

Figure 3 Left: Image reconstruction of the RSG V602 Car based on VLTI-PIONIER data. Right: The most similar image from a series of 3D convection snapshots convolved to the resolution of the observation. From Climent et al., in prep.

These images are mostly obtained at a single or very few epochs, although both AGB and RSG stars are long-period variables. Much stronger constraints of dynamic models can be obtained by time-series observations covering intra-cycle and inter-cycle variability. First such observations in the continuum and in molecular bands using the GRAVITY instrument (Fig. 4) have been presented by Wittkowski et al. (2018). Eventually, time-series of imaging will be needed, as the features are morphologically complex and cannot easily be described by geometrical models. Like-wise imaging at higher spectral resolution but still at single epochs are succeeding as well (e.g. Ohnaka et al. 2017).

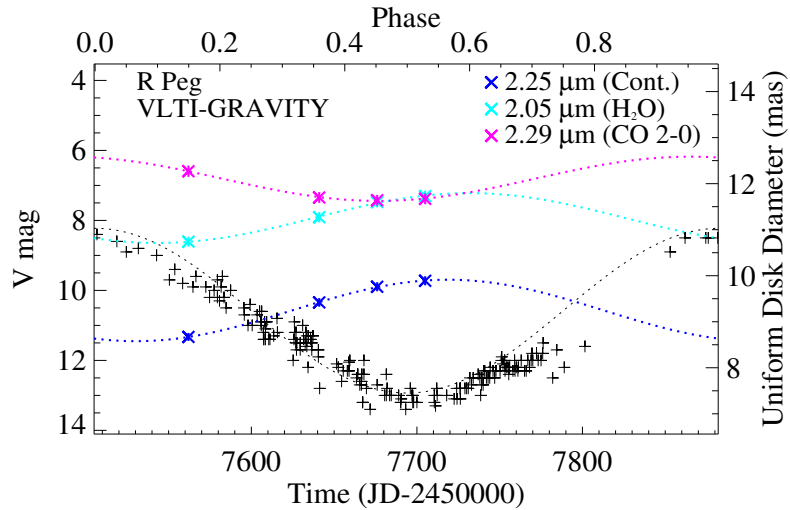

Figure 4. Variability of the Mira R Peg in V mag and in the UD angular diameters in the near-continuum band, and in bands dominated by H2O (2.05 μm) and CO 2–0 (2.29 μm). The minimum continuum size tracks the maximum light. The minimum contribution of H2O tracks the maximum light as well. The contribution from CO instead is greatest at maximum light, indicating a different and more stable behavior of CO compared to water vapor. From Wittkowski et al. (2018).

### Desirable Improvements for the 2030s

The VLTI will continue to provide the highest spatial resolution at ESO in the era of the ELT. Only the apparently largest stars can be resolved by the ELT, or by ALMA, while we need to observe a representative sample of different stellar parameters, for which long-baseline interferometry is mandatory.

### Instrumental Improvements

- Shorter (near-IR *J*-band or visible) wavelengths: The advantage of instruments at shorter wavelengths is two-fold: (1) The absolute spatial resolution is larger, and (2) the contrast of surface features is larger at shorter wavelengths (e.g. Patru et al. 2010, Chiavassa et al. 2011). AGB and RSG stars are red targets and faint in the visible; The near-IR *J*-band may be a good compromise for these targets.
- Longer baselines: The availability of the longest baselines available on the Paranal platform (by doubling delay lines) will provide a higher spatial resolution.
- Combination of more beams: A well-filled *uv* plane is important for an image reconstruction independent of prior images. Instruments combining 6 or 8 beams help.
- Higher contrast: Both AGB and RSG stars show relatively faint atmospheric layers that cannot be appreciated with the currently achieved dynamic range of reconstructed images. Instruments that provide higher visibility accuracies are needed.
- Spectral resolution: A high spectral resolution (>10000) is needed for velocity-resolved imaging.

### Operational Improvements

- More flexible relocations of ATs: Imaging programs need a minimum number of observations on several different telescope configurations to be completed. VLTI Operations are already moving to a more flexible way of relocations to make sure that the required number of observations are taken before moving to the next configuration. Such efforts need to be expanded for next-generation scheduling tools.
- More (or only) service mode: A larger fraction of service mode is favorable for a more flexible medium-term scheduling of AT configurations, taking losses into account for the time of relocations. At the time of the ELT, a pure service operations scheme for the VLTI should be considered.
- Enabling time series of imaging: The schedule and its tools needs to take into account requirements for time-series of imaging, while also considering requirements for other monitoring campaigns with different cadences.

## Synergies in the 2030s

The ELT and ALMA can resolve the apparently largest stars at higher imaging fidelity. For a larger representative sample of stars, optical interferometry continuous to be mandatory in the era of the ELT; The ELT and ALMA can complement imaging of circumstellar environments and in lines of dust-forming molecules. Other interferometric facilities (e.g. CHARA) provide larger baselines, but not in the southern hemisphere, and may not provide other instrumental and operational aspects outlined above. For targets around the equator *uv* points may be shared and combined.

## References

Airapetian V., Carpenter K. G., Ofman L., 2010, ApJ, 723, 1210
Arroyo-Torres B., et al., 2015, A&A, 575, A50
Chiavassa A., et al., 2011, A&A, 528, A120
Climent, J. B. et al., in prep.
Freytag B., Hoefner S., 2008, A&A, 483, 571
Freytag B., Liljegren S., Hoefner S., 2017, A&A, 600, A137
Hoefner S., Olofsson H., 2018, A&ARv, 26, 1
Josselin E., Plez B., 2007, A&A, 469, 671
Montarges M., et al., 2016, A&A, 588, A130
Ohnaka K., Weigelt G., Hofmann K.-H., 2017, Nature, 548, 310
Patru, F., Chiavassa, A., Mourard, D., Tarmoul, N. 2010, Proc. SPIE, 7734, 77341G
Thirumalai A., Heyl J. S., 2012, MNRAS, 422, 1272
Vlemmings W., et al., 2017, Nature Astronomy, 1, 848
Vlemmings W. H. T., et al., 2018, A&A, 613, L4
Wittkowski M., et al., 2016, A&A, 587, A12
Wittkowski M., et al., 2017, A&A, 606, L1
Wittkowski M., et al., 2017a, A&A, 601, A3
Wittkowski M., et al., 2018, A&A, 613, L7